\documentclass[useAMS,usenatbib]{mn2e}
\usepackage{epsfig}
\usepackage{amsbsy}
\usepackage{hyperref}

\providecommand{\eprint}[1]{\href{http://arxiv.org/abs/#1}{#1}}
\providecommand{\adsurl}[1]{\href{#1}{ADS}}

\newcommand{\lya}{Lyman-$\alpha$~}
\newcommand{\gad} {{\small {GADGET-2}}\,}

\newcommand{\be}{\begin{equation}}
\newcommand{\ee}{\end{equation}}
\newcommand{\ba}{\begin{eqnarray}}
\newcommand{\ea}{\end{eqnarray}}
\newcommand{\brr}{\begin{array}}

\newcommand{\err}{\end{array}}
\newcommand{\bc}{\begin{center}}
\newcommand{\ec}{\end{center}}

\newcommand{\ns}{{\mbox{{\it n}$_{\rm s}$}}}

\newcommand{\nrun}{{\mbox{n$_{\rm run}$}}}

\newcommand{\apj}{\mbox ApJ}

\DeclareMathAlphabet{\mathsc}{OT1}{cmr}{m}{sc}
\def\testbx{bx}%
\DeclareRobustCommand{\ion}[2]{%
\relax\ifmmode
\ifx\testbx\f@series
{\mathbf{#1\,\mathsc{#2}}}\else
{\mathrm{#1\,\mathsc{#2}}}\fi
\else\textup{#1\,{\mdseries\textsc{#2}}}%
\fi}
\title[The \lya forest and WMAP year three]
{The \lya forest and WMAP year three}

\author[Matteo Viel, Martin G. Haehnelt \& Antony  Lewis]
{Matteo  Viel$^{1,2}$, Martin G. Haehnelt$^{1}$ \& Antony Lewis$^{1}$\\ $^1$ Institute of
Astronomy, Madingley Road, Cambridge CB3 0HA\\
$^2$ INAF - Osservatorio Astronomico di Trieste, Via G.B. Tiepolo 11,
I-34131 Trieste, Italy
\\}

\begin{document}
\maketitle
\begin{abstract}
A combined analysis of Cosmic Microwave Background (CMB) and \lya
forest data can constrain the matter power spectrum from small
scales of about $1h^{-1}$ Mpc all the way to the horizon scale. The
long lever arm and complementarity provided by such an analysis has
previously led to a significant tightening of the constraints on the
shape and the amplitude of the power spectrum of primordial density
fluctuations. We present here a combined analysis of the WMAP three
year results with \lya forest data. The amplitude of the matter power
spectrum $\sigma_{8}$ and the spectral index $\ns$ inferred from the
joint analysis of high and low resolution 
\lya forest data as analysed by \citet{Viel:2005ha} are consistent with
the new WMAP results to within $1\sigma$. The joint analysis with the
mainly low resolution data as analysed by \citet{McDonald:2004xn} suggest a
value of $\sigma_8$ that is $\sim 2\sigma$ higher than that inferred
from the WMAP three year data alone. The joint analysis of the three
year WMAP and the \lya forest data also does not favour a running of
the spectral index.  The best fit values for a combined analysis of the
three year WMAP data, other CMB data, 2dF and the \lya forest data are
$(\sigma_8,\ns) = (0.78\pm 0.03,0.96 \pm 0.01$).
\end{abstract}


\section{Introduction}
Measurements of the matter power spectrum from \lya forest data extend
to smaller scales and probe a redshift range complementary to
estimates of the matter power spectrum from Cosmic Microwave
Background (CMB), galaxy surveys or weak gravitational lensing
observations (e.g. \citet{Croft:1997jf}, \citet{Gnedin:1997td}, \citet{McDonald:1999dt}, \citet{Hui:2000rw}, \citet[C02]{Croft:2000hs}, \citet{McDonald:2001fe}, \citet{Viel:2002gn}, \citet{Meiksin:2003qb}, \citet[VHS]{Viel:2004bf}).

The combined analysis of \lya forest data with the first year of WMAP
data \citep[WMAP1]{Spergel:2003cb} suggested that the fluctuation
amplitude of the matter power spectrum on small scales was rather high
($\sigma_{8} \sim 0.9$) and that there was no significant deviation of
the spectral index of primordial density fluctuations from a
Harrison-Zeldovich spectrum ($\ns=1$). There was also no evidence for a
(large) running of the spectral index, a non-zero neutrino mass or a
deviation from a cold dark matter spectrum at small scales (\citet[VHS]{Viel:2004bf},  \citet{Viel:2004np}, \citet[M05]{McDonald:2004xn}, \citet{Viel:2005qj}, \citet{Seljak:2004xh}, \citet{Lidz:2005gz}, \citet{Beltran:2005gr}, \citet{Abazajian:2005xn}).

\begin{figure*}
\setlength{\unitlength}{1cm} \centering
\begin{picture}(19,9)
\put(0, 0){\includegraphics{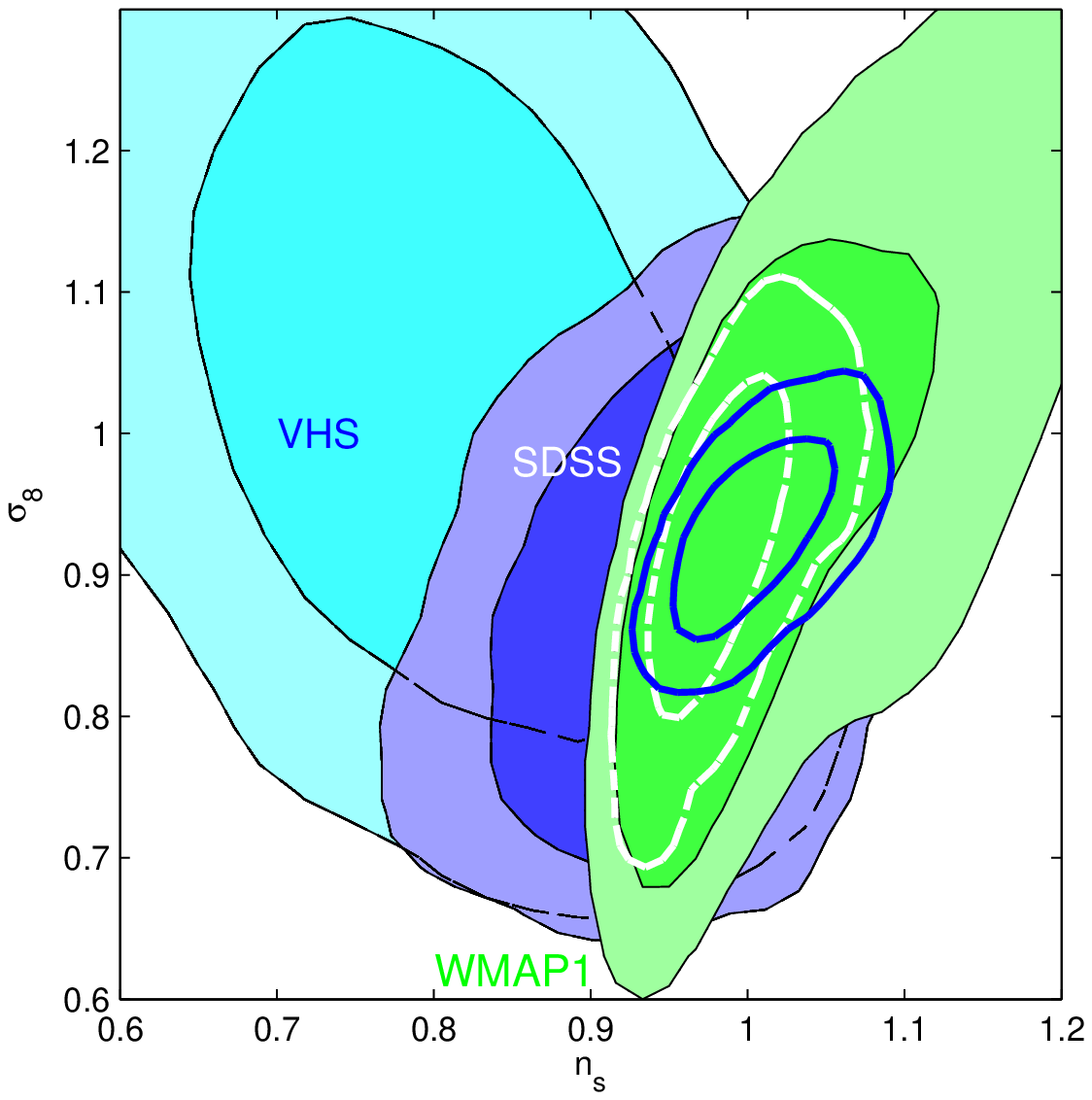}}
\put(8.5, 0){\includegraphics{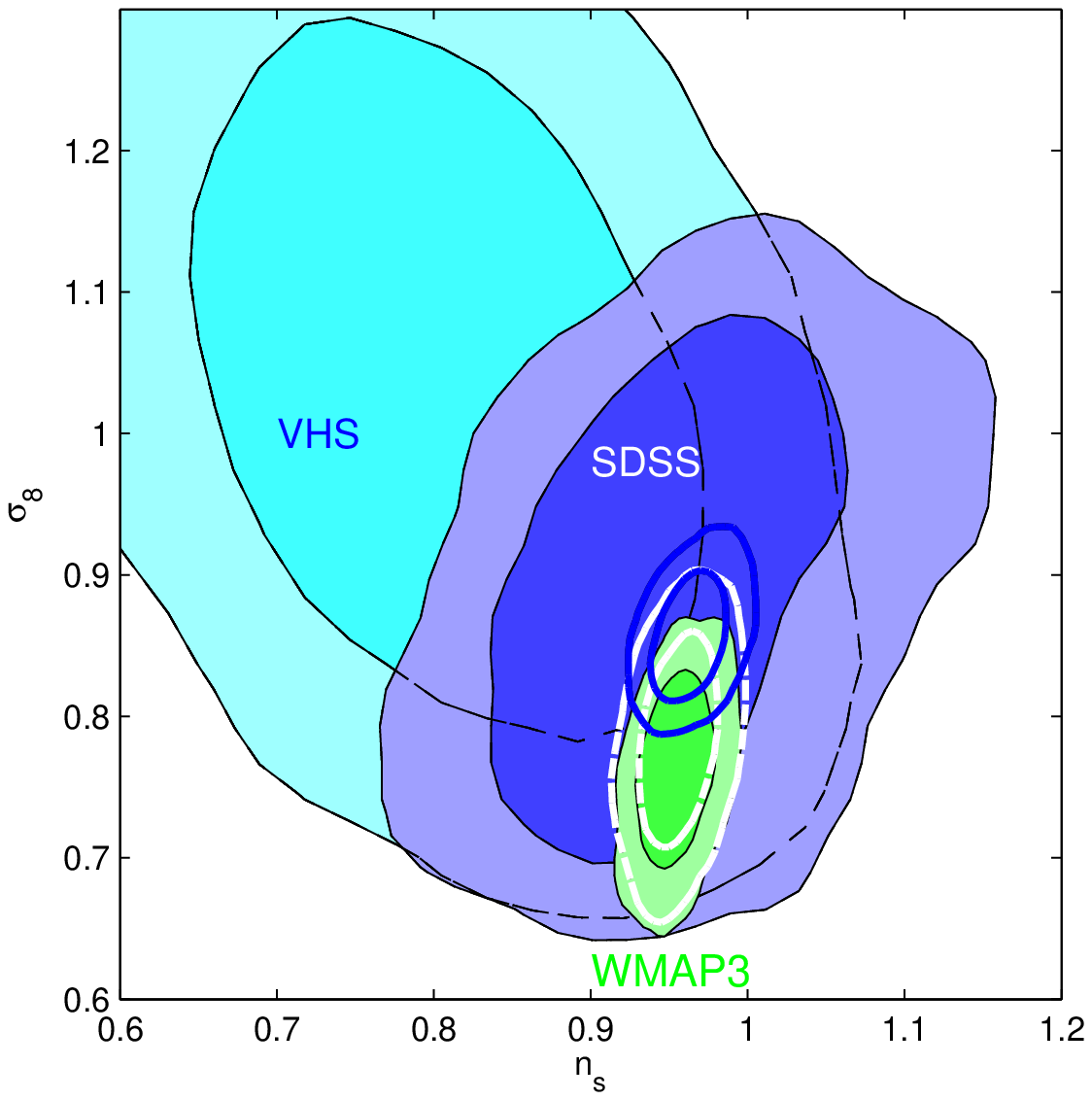}}
\end{picture}
\caption{$1$ and $2\sigma$ likelihoods for $\sigma_8$ and $\ns$
marginalized over all  other parameters.  {\it Left panel:}
Constraints are for WMAP1 only (green),the LUQAS+CROFT data sets as
analysed by VHS (cyan) and the SDSS \lya forest data of M05 (blue).
The thick dashed white empty contours refer to WMAP1 + VHS, while the
solid blue contours are for WMAP1 + SDSS. {\it Right panel:} As in the
left panel but for the WMAP3 data set. }

\label{fig1}
\end{figure*}
\citet{Viel:2004np} found $\sigma_8=0.94\pm0.08$, $n=0.99\pm0.03$
($1\sigma$) and no evidence for a (large) running of the spectral in a
combined analysis of a large sample of high resolution spectra quasar
(QSO) absorption spectra at $z\sim 2.5$~\citep{Kim:2003qt,Croft:2000hs}
and the WMAP1 data.  Similar results, with somewhat smaller errors
($\sigma_8=0.90\pm 0.03$, $\ns=0.98\pm0.02$) have been subsequently
obtained by the SDSS collaboration in a combined analysis of the WMAP1
and other CMB data, SDSS galaxy survey data and SDSS \lya forest data
\citep{Seljak:2004xh}.  The \lya forest data analysed by M05 and
\citet{Seljak:2004xh} consists mainly of low-resolution low S/N
SDSS spectra with a wide redshift coverage ($2<z<4$) to which they
added a small sample of eight high resolution spectra
\citep{McDonald:1999dt}. The flux power spectrum was modelled using
dark matter simulations, which take into account hydrodynamical effects in an
approximate way and were calibrated with a few hydrodynamical
simulations.
\citet{Viel:2005ha} found $\sigma_8=0.91\pm0.07$, $\ns=0.95\pm0.04$ for
the SDSS \lya forest data alone using a suite of state-of-the-art full
hydrodynamical simulations.  Further studies of \lya forest data by
\citet{Desjacques:2004xy}, \citet{Jena:2004fc} and
\citet{Zaroubi:2005xx} also came to similar conclusions.

The WMAP3 data alone argues now for significant deviation from a
Harrison-Zeldovich spectrum, $\ns=0.95 \pm 0.02$, and a smaller value
for the fluctuation amplitude on small scales $\sigma_{8} = 0.74\pm
0.06$ \citep{Spergel:2006hy}. The WMAP team chose not to update their
combined analysis of CMB and \lya forest in their WMAP3 data release.
In this Letter we will present such a joint analysis.

\section{The data sets}

\subsection{WMAP}
The WMAP\footnote{\url{http://lambda.gfsc.nasa.gov}} satellite has mapped
the entire sky in five frequency bands between 23 and 94 GHz with
polarization sensitive radiometers.  The temperature power spectrum has been
measured over a large range of scales ($l<1000$) to an
unprecedented accuracy \citep{Hinshaw:2006ia}. We
will use the temperature and polarization \citep{Page:2006hz} power spectra and maps
as used by the WMAP likelihood codes \citep{Verde:2003ey, Spergel:2006hy} as implemented in the code
COSMOMC\footnote{\url{http://cosmologist.info/cosmomc/}}~\citep{Lewis:2002ah}.

\begin{figure*}
\setlength{\unitlength} {1cm} \centering
\begin{picture}(19,9)
\put(0, 0){\includegraphics{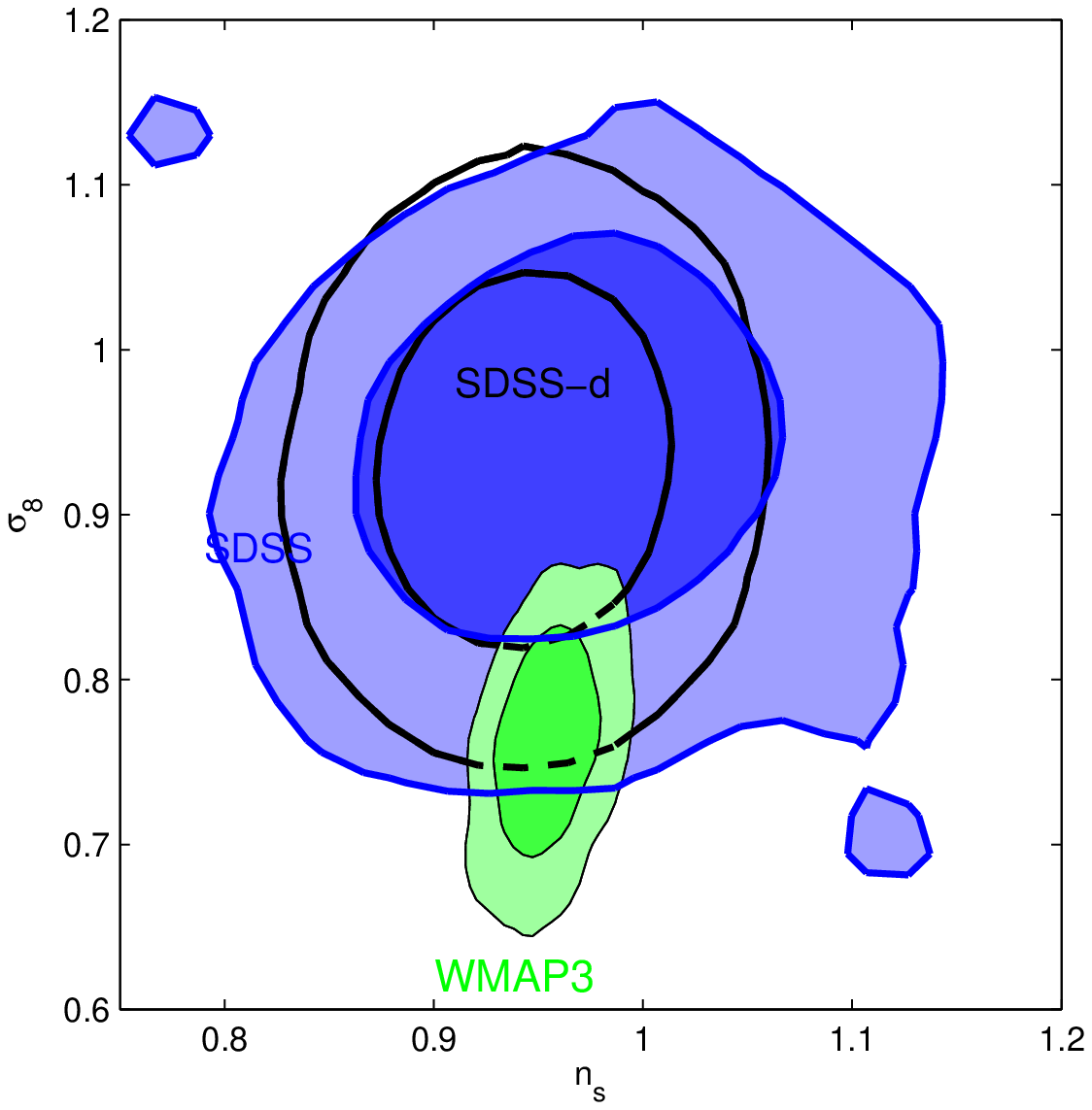}}
\put(8.5, 0){\includegraphics{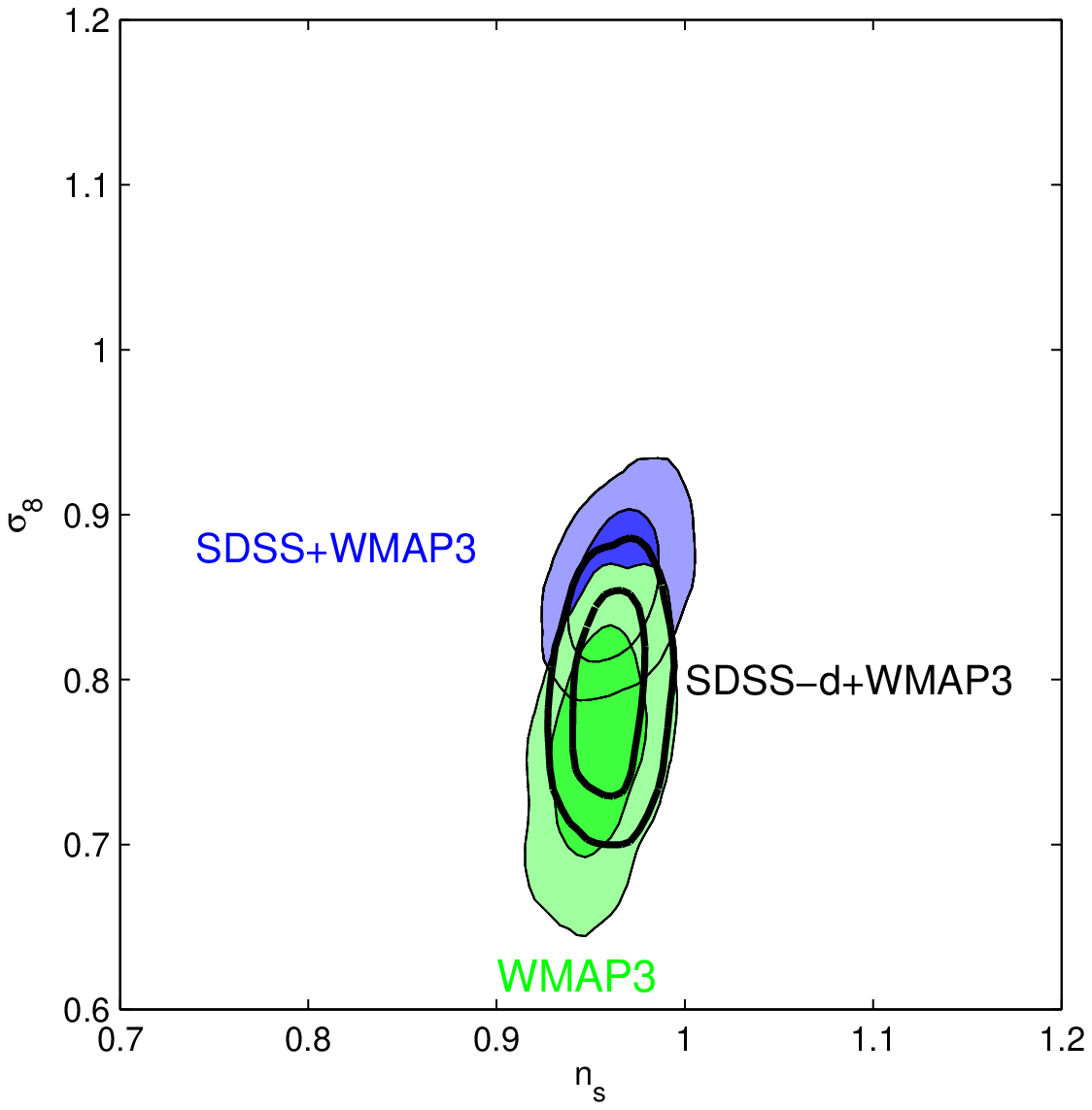}}
\end{picture}
\caption{$1$- and $2\sigma$ likelihoods for $\sigma_8$ and $\ns$
marginalized over all  other parameters.  {\it Left panel:}
The blue contours  show the constraints  for SDSS only as
analysed by M05 with an HST prior $72\pm 8$ km/s/Mpc
(blue). The empty solid contours are for
the SDSS data set as analysed by \citet[SDSS-d]{Viel:2005ha}.
The WMAP3 results are shown in green. {\it Right panel:} The combined analysis with WMAP3
using the same colour/line coding as in the left panel.}
\label{fig2}
\end{figure*}

\subsection{\lya forest data sets}

We will investigate two different \lya forest data sets.  The sample
of high resolution QSO absorption spectra used by VHS and
\citet{Viel:2004np}, consisting of the LUQAS sample (27 high resolution
QSOs) \citep{Kim:2003qt}
and the (reanalysed) sample of C02 (30 high resolution and 23 low
resolution spectra), and the SDSS \lya forest data as
presented by \citet{McDonald:2004eu}.  The SDSS \lya forest data set
consists of $3035$ QSO spectra with low resolution ($R\sim 2000$) and
low S/N ($\sim 10$ per pixel) spanning a wide range of redshifts
($z=2.2-4.2$), while the LUQAS and the C02 samples contain mainly high
resolution ($R \sim 45000$), high signal-to-noise ($>50$ per pixel)
QSO spectra with median redshifts of $z=2.125$ and $z=2.72$,
respectively.  Modelling the flux power spectrum of the \lya forest
accurately for given cosmological parameters is not as straightforward
as modelling the CMB power spectra and accurate numerical simulations
are required.  M05 modelled the flux power spectrum using a large
number of Hydro Particle Mesh (HPM) simulations~\citep{Gnedin:1997td,Viel:2005eg},
calibrated with a few full
hydrodynamical simulations.
VHS improved instead the effective bias method
developed by C02 (see~\citet{Gnedin:2001wg} and \citet{Zaldarriaga:2001xs} for a critical assessment of the errors involved), by using a
grid of full hydrodynamical simulations run with the Tree-SPH code
\gad \citep{Springel:2000yr, Springel:2005mi} to infer the
linear matter power spectrum. \citet{Viel:2005ha} used a Taylor
expansion of the flux power spectrum around best fitting values based
on full hydrodynamical simulations to model the dependence of the flux
power on cosmological and astrophysical parameters in their
independent analysis of the SDSS \lya forest data.

\begin{table*}
\begin{center}
\caption{Summary of the constraints on $\sigma_8$ and $\ns$ for the VHS
\citep{Viel:2004bf} and SDSS~\citep{McDonald:2004xn} samples. SDSS-d refers to the analysis of SDSS data by \citet{Viel:2005ha}. The
quoted errors are the 68\%
confidence limits.}
\label{tab:params}
\begin{tabular}{lccccccc}
\hline
& \tiny{WMAP1} & \tiny{WMAP1+VHS} & \tiny{WMAP1+SDSS} & \tiny{WMAP3} & \tiny{WMAP3+VHS} & \tiny{WMAP3+SDSS} & \tiny{WMAP3+SDSS-d} \\
\hline
$\sigma_8$  & $0.90\pm0.10$  & $0.94 \pm 0.08$ & $0.93\pm0.04$  &  $0.76 \pm 0.05$ & $0.78 \pm 0.05$ & $0.86\pm0.03$ & $0.80\pm0.04 $  \\
$\ns$  & $0.99\pm0.04$ & $0.99\pm 0.03$  & $ 0.99\pm0.03$ & $0.96\pm 0.02$&  $0.96 \pm 0.02$ & $0.96\pm0.02$ & $0.96 \pm 0.01$ \\
\hline
\end{tabular}
\end{center}
\end{table*}

\section{Results}
\label{results}

\subsection{Incorporating the \lya  data into COSMOMC}

The linear dark matter power spectra inferred from the two \lya forest
data sets have been incorporated into the new public available version of
COSMOMC.  The VHS power spectrum consists of estimates of the linear
dark matter power spectrum at nine values of the wavenumber $k$  at $z=2.125$
and nine values at $z=2.72$, in the range
$0.003<k$ (s/km)$<0.03$.  The estimate of the uncertainty of
the overall amplitude of the matter power spectrum is  $29\%$. This
estimate takes into account possible systematic
and statistical errors (see the relevant tables of VHS for a detailed
discussion).  M05 provided a measurement of slope and amplitude of the
matter power spectrum at $z=3$  at a wavenumber $k=0.009$ s/km with an
estimate of the $1\sigma$ error of the amplitude of $\sim 14\%$.  M05
have also made available a table that gives the
minimum $\chi^2$ for a given cosmological model as a function of the
amplitude and slope after marginalization over a wide range of
cosmological and ``nuisance'' parameters.  The
nuisance parameters characterize a range of astrophysical and
noise-related systematic uncertainties. We have furthermore
implemented the modelling of the SDSS flux power spectrum with the
method of \citet{Viel:2005ha} based on a Taylor expansion of the
flux power spectrum around a best fitting model.

\subsection{Constraints on $\sigma_{8}$ and $\ns$}

To make contact with previous analyses we show the marginalized
$1\sigma$ and $2\sigma$ likelihoods in the $\sigma_{8}$-$\ns$
projection in Figure 1. The coloured contours in the left panel show
the constraints for the VHS sample (light cyan), the SDSS sample (dark
blue) with the likelihood estimates provided by M05 and the
constraints for the WMAP1 data (light green). In all cases we assume
the Universe to be flat, no  contribution from tensor perturbations,
a pure cosmological constant ($w=-1$) and 
neutrinos with negligible mass.  For the analysis of the WMAP1 data
we assumed a prior on the Thomson optical depth $0<\tau<0.3$.

The dark solid (SDSS) and light dashed (VHS) contours show the joint
constraints for the \lya forest and WMAP1 data. As pointed out by
\citet{Viel:2005ha}, there is remarkable agreement between the two
joint analyses with the high resolution absorption spectra as analysed
by VHS and \citet{Viel:2004np} and the larger sample of low resolution
SDSS spectra as analysed by M05 and \citet{Seljak:2004xh}.  The \lya
forest data break some of the degeneracies of the WMAP1 data which are
responsible for the elongated shape of the error contours.  The joint
analysis tightens the constraints in the $\sigma_{8}-\ns$ plane by a
factor $\sim 2$ (VHS) and $\sim 4$ (M05), compared to the constraints
from the WMAP1 data alone but offers less help in improving the
constraints on $\ns$.  As discussed above and summarized in Table 1, in
the combined analysis with the WMAP1 data the best fitting value of the
spectral index is not significantly different from $\ns=1$ and
$\sigma_8 \sim 0.9$.  The significantly smaller error bars of the SDSS
data set are due to the much larger sample: the wider range of
redshifts covered is responsible for breaking some of the degeneracies
intrinsic to the \lya forest data (M05).

In the right panel of Figure \ref{fig1} we show how the situation has
changed with the WMAP3 data. As discussed in \citet{Spergel:2006hy} and
\citet{Page:2006hz}, the contours for the WMAP3 data have shrunk by a
factor $\sim 3$ compared to those of WMAP1 and lie at the bottom part
of the region allowed by the latter.  This is mainly due to the
improved measurement of the optical depth from the large scale polarization
\citep{Page:2006hz,Lewis:2006ma}.

In the case of the high resolution VHS sample the errors are too large
to significantly tighten the WMAP3 constraints. The joint analysis of
WMAP3 with the SDSS \lya forest data places now even tighter
constraints at the bottom end of the range preferred by the SDSS \lya
forest data alone, about $2\sigma$ above the best fit value from the
WMAP3 only data: $\sigma_8=0.86 \pm 0.03$ (see
also~\citet{Lewis:2006ma}).  The discrepancy is larger than may be
naively inferred from the overlap of the \lya only and WMAP3 only
analysis because  the data sets prefer different values for some of the other
parameters in particular $\Omega_{\rm m}$.  The best fit value for
$\Omega_{\rm m}$ $(\sim 0.3$) is also $\sim 2\sigma$ higher for the
combined analysis than for the WMAP3 data alone.
The SDSS \lya forest data as analysed by \citet{McDonald:2004xn}
and the new WMAP results appear to be marginally consistent.  Possible
explanations for the (small) discrepancy, if there exists one, may be
somewhat too optimistic errors for one or both of the data sets.  As
discussed in \citet{Page:2006hz} the polarization measurements of the
CMB are very difficult mainly due to foreground
polarization. Moreover, modelling the \lya forest data also has a
range of not yet fully understood systematic uncertainties.

As discussed extensively by \citep{McDonald:2004xp} \citet{Viel:2005ha} the mayor
systematic uncertainties are the still not very accurately known
thermal state of the IGM, the determination of the effective optical
depth, the modelling of the effect of strong
absorption line systems (\citet{Viel:2004st}) and UV fluctuations
and the remaining deficiencies in our ability to accurately predict
the flux power spectrum for a large parameter space.
To investigate the last issue issue further we compare in the left panel of
Figure 2 the analysis of the SDSS data by M05 with that by \citet{Viel:2005ha}
which is based on a Taylor expansion of the flux power
spectrum around a best fitting model (labelled as SDSS-d in Figure
2). The analysis of \citet{Viel:2005ha} uses more accurate full
hydro-simulations instead of the approximate simulations of M05 at the
expense of a much less complete sampling of parameter space,
especially far from the best-fitting values, where the errors are
possibly underestimated.
Note that here, both for the forest data and
the joint analysis, we assumed an HST prior for the Hubble
constant~\citep{Freedman:2000cf} that significantly shrinks the error contours
for the \lya forest data compared to that of Figure 1. As discussed by
\citet{Viel:2005ha} there is remarkable agreement between the two
analyses (note that the analysis of \citet{Viel:2005ha} does not use
the last three redshift bins nor high resolution QSOs compared to that
of M05). In the right panel of Figure 2 we show the constraints for
the joint analysis with the WMAP3 data. The SDSS \lya forest data as
analysed by \citet{Viel:2005ha} combined with WMAP3 give a smaller
value best fit value of $\sigma_8=0.80\pm0.04$ which is in agreement with
that from the WMAP3 data alone to within $1\sigma$.  The joint
analysis of \lya forest data and the new WMAP data including a
possible running of the spectral index gives $\nrun = -0.002 \pm
0.015$ at $k=0.002$ Mpc$^{-1}$ also in agreement with the estimate by
\citet{Spergel:2006hy}.

We have also performed an extended combined analysis that includes
the further CMB experiments  ACBAR~\citep{Kuo:2002ua}, CBI
\citep{Readhead:2004gy}, VSA~\citep{Dickinson:2004yr} , the 2dF galaxy
power spectrum \citep{Percival:2001hw} and the VHS and SDSS-d
\lya forest data. In this case we get
$(\sigma_8,\ns) = (0.78\pm 0.03,0.96 \pm 0.01)$.
Further results are listed in Table \ref{tab:params2}.

\begin{table*}
\begin{center}
\caption{The marginalized constraints on cosmological parameters from WMAP3 and other data sets.
VHS refers to the LUQAS+CROFT sample as analysed by \citet{Viel:2004bf}; SDSS refers to the  measurement by \citet{McDonald:2004xn}; EXT refers to smaller scales CMB data sets: ACBAR~\citep{Kuo:2002ua}, CBI~\citep{Readhead:2004gy}, VSA~\citep{Dickinson:2004yr}
and the 2dF galaxy survey \citep{Percival:2001hw}. SDSS-d
refers to the SDSS analysis  by  \citet{Viel:2005ha}.}
\label{tab:params2}
\begin{tabular}{lcccccc}
\hline
&  \tiny{WMAP3} & \tiny{WMAP3+VHS} & \tiny{WMAP3+SDSS} & \tiny{WMAP3+SDSS-d} & \tiny{WMAP3+EXT+SDSS-d} & \tiny{WMAP3+SDSS-d(runn.)}\\
\hline
$\Omega_{\rm c}h^2$ & $0.106 \pm 0.007$ & $0.109 \pm 0.008$ & $0.120\pm0.006$ & $0.110 \pm 0.006$ & $0.109 \pm 0.006$  & $0.112 \pm 0.007$ \\
$10^2\Omega_{\rm b}h^2$ & $2.222\pm 0.069$  & $2.237\pm 0.072$  & $2.277\pm0.065$  & $2.226 \pm 0.071$&  $2.224 \pm 0.066 $ & $2.221 \pm 0.094$ \\
$\Omega_{\rm m}$ & $0.242\pm 0.032$  & $0.257\pm 0.037$  & $0.304\pm0.031$ & $0.258\pm0.029$ & $0.253 \pm 0.028$ & $0.269 \pm 0.036$\\
$h$  & $0.729 \pm 0.029$ & $0.719 \pm 0.031$ & $0.688\pm0.025$ & $0.719\pm0.026$ & $0.723 \pm 0.025$ & $0.711 \pm 0.034$\\
$\tau$ & $0.089\pm 0.030$  & $0.092\pm 0.029$  & $0.101\pm0.028$& $0.098\pm0.032$  & $0.104\pm  0.036$& $0.104 \pm 0.030$ \\
$\sigma_8$ & $0.761 \pm 0.046$ & $0.784 \pm 0.048$ & $0.857\pm0.028$ & $0.801\pm0.039$ & $0.785 \pm 0.035$ & $0.800 \pm 0.037$ \\
$\ns$ & $0.956 \pm 0.016$ & $0.956\pm 0.017$  & $0.964\pm0.016$  & $0.960\pm0.013$ & $0.957 \pm 0.014$ & $0.963 \pm 0.020$\\
$n_{\rm{run}}$& & - & - & - & - & $-0.002 \pm 0.015$\\
\hline
\end{tabular}
\end{center}
\end{table*}

\section{Conclusions}
We have performed a combined analysis of the WMAP three year  results with
high and low resolution \lya forest data in order to
constrain the shape of the power spectrum of primordial density
fluctuations and the amplitude of the matter power spectrum at
intermediate scales $\sigma_{8}$.  The main results are as follows.

\begin{itemize}

\item
The high resolution VHS \lya forest data is consistent to within
$1\sigma$ with the three year WMAP results but offers little additional
constraining power due to the large error bars.  The larger sample of
mainly low resolution \lya forest data (SDSS) as analysed by
\citet{Viel:2005ha} is also consistent to within $1\sigma$ with the new
WMAP results. However the joint analysis of the SDSS data gives about
$2\sigma$ higher $\sigma_8$ and $\Omega_{\rm m}$ values than those inferred
from the new WMAP results alone.

\item
The best fit values for a combined analysis are ($1\sigma$):
$(\sigma_8,\ns) = (0.78\pm 0.05,0.96 \pm 0.02)$ and $(\sigma_8,\ns) =
(0.86\pm 0.03,0.96 \pm 0.02)$ for WMAP combined with high resolution
\lya forest data and WMAP combined with low resolution \lya forest
data as analysed by M05.
The analysis of the SDSS data set as analysed by \citet{Viel:2005ha}
based on full  hydrodynamical simulations gives $\sigma_8 = 0.80\pm 0.04$.

\item
The joint analysis of \lya forest data and the new WMAP data
does not favour a running of the spectral index. The best fitting
value is $\nrun = -0.002 \pm 0.015$ at $k=0.002$
Mpc$^{-1}$.

\item
Adding other CMB data sets, the 2dF galaxy survey and both
\lya data sets the constraints on the matter power spectrum
become $(\sigma_8,\ns) = (0.78\pm 0.03,0.96 \pm 0.01$).
\end{itemize}

The \lya forest data appears to be in reasonable agreement
with the CMB and other data sets which probe the matter power
spectrum at larger scales. The \lya forest data will thus
continue to unfold its special power to measure parameters
that affect the overall shape and/or the small scale part of the
matter power spectrum.  For the near future further progress is,
however, likely to be driven by a better understanding of the
systematic uncertainties, rather than the compilation of larger
data sets.

\section*{Acknowledgments.}
Numerical computations were done on the COSMOS supercomputer at DAMTP
in Cambridge. COSMOS is a UK-CCC facility which is supported by HEFCE
and PPARC. We acknowledge the use of the Legacy Archive for Microwave
Background Data Analysis (LAMBDA). Support for LAMBDA is provided by
the NASA office of Space Science. We thank K. Abazajian and J.
Lesgourgues for providing some of the \lya modules for the likelihood
analysis and for useful suggestions. AL is supported by a PPARC
Advanced Fellowship.  We thank Joop Schaye for a useful referee report.


\providecommand{\aj}{ApJ }\providecommand{\apj}{ApJ
  }\providecommand{\apjl}{ApJ }\providecommand{\mnras}{MNRAS}

\end{document}